\documentclass[citeautoscript,floatfix,aps,prb,twocolumn,superscriptaddress]{revtex4-1}
\usepackage[utf8]{inputenc}
\usepackage{graphicx}
\usepackage{epstopdf}
\usepackage{natbib}
\usepackage{array}
\usepackage{braket}
\usepackage{float}
\usepackage{fancybox}
\usepackage[usenames,dvipsnames]{color}
\usepackage{dcolumn}
\usepackage{bm}
\usepackage{soul}  
\usepackage{changes}
\usepackage{hyperref}
\usepackage[T1]{fontenc}
\usepackage{xcolor}
\usepackage{upgreek}
\newcommand{\vb}{V$_{\text{B}}^{-}$}
\newcommand{\nv}{NV$^{-}$}

\begin{document}

\title{Two-dimensional materials as a multiproperty sensing platform}

\author
{Dipankar Jana$^{1,\#}$, Shubhrasish Mukherjee$^{1,\#}$, Dmitrii Litvinov$^{1,2,\#}$, Magdalena~Grzeszczyk$^{1,\#}$, Sergey Grebenchuk$^{1,\#}$, Makars~\v{S}i\v{s}kins$^{1,\#}$, Virgil Gavriliuc$^{1,\#}$, Yihang Ouyang$^{1,2}$, Changyi Chen$^{1,2}$, Yuxuan Ye$^{1,2}$, Yiming Meng$^{1,2}$, Maciej Koperski$^{*,1,2}$
\\
\ \\
\normalsize{$^1$~Institute for Functional Intelligent Materials, National University of Singapore, 117544, Singapore}\\
\normalsize{$^2$~Department of Materials Science and Engineering, National University of Singapore, 117575, Singapore}\\
\normalsize{$^\#$~These authors contributed equally to this work.}\\
\ \\
\normalsize{*Correspondences to: msemaci@nus.edu.sg}\\ 
}

\begin{abstract}
Two-dimensional (2D) materials have disrupted materials science due to the development of van der Waals technology. It enables the stacking of ultrathin layers of materials characterized by vastly different electronic structures to create man-made heterostructures and devices with rationally tailored properties, circumventing limitations of matching crystal structures, lattice constants, and geometry of constituent materials and supporting substrates. 2D materials exhibit extraordinary mechanical flexibility, strong light-matter interactions driven by their excitonic response, single photon emission from atomic centers,  stable ferromagnetism in sub-nm thin films, fractional quantum Hall effect in high-quality devices, and chemoselectivity at ultrahigh surface-to-volume ratio. Consequently, van der Waals heterostructures with atomically flat interfaces demonstrate an unprecedented degree of intertwined mechanical, chemical, optoelectronic, and magnetic properties. This constitutes a foundation for multiproperty sensing, based on complex intra- and intermaterial interactions, and a robust response to external stimuli originating from the environment. Here, we review recent progress in the development of sensing applications with 2D materials, highlighting the areas where van der Waals heterostructures offer the highest sensitivity, simultaneous responses to multiple distinct externalities due to their atomic thickness in conjunction with unique material combinations, and conceptually new sensing methodology.
\end{abstract}

\maketitle
\section*{Introduction}

Sensing applications rely on mutual relationships between physical quantities occurring within a singular material, between several distinct materials, or under the influence of the environment. From this perspective, the scope of sensing is quite broad, encompassing intertwined mechanical, chemical, optoelectronic, and magnetic properties. Two-dimensional (2D) materials offer unique capabilities in the domain of sensing, as their fabrication technology has been developed to combine vastly different materials separated only by a van der Waals gap \cite{Geim2013, Novoselov2016, Liu2016, Castellanos-Gomez2022, Liu2024}. This favors strong interactions at the interfaces, enabling on-demand and rationally engineered sensing functionalities that are unlocked via specific combinations of materials. 

Monolayers of 2D materials are typically characterized by sub-nm thicknesses, which contribute to their sensing capabilities. In such atomically thin layers, electronic and spin states exist essentially at the surface, making them particularly sensitive to the environment. Robust environmental sensitivity is exemplified by 2D-confined excitons in a semiconductor sensing the fractional quantum Hall effect in a proximitized semimetal via modulations of an effective dielectric constant, or by the emergence of ultrahigh proximity magnetic fields at semiconductor-ferromagnet interfaces. The large surface-to-volume ratio of atomically thin layers ensures optimal sensitivity achieved at a minimal mass of the material, which is particularly evident in chemosensing applications. The detection of gas molecules or DNA sequencing can be realized at low material loading, which is beneficial for biocompatible and sustainable 2D sensing technology. 

Moreover, 2D materials may be deposited on arbitrary substrates, including the fabrication of suspending layers. Their extraordinary flexibility has been utilized to create nanodrum membranes capable of detecting structural, electronic, and magnetic phase transitions as well as minuscule external stimuli from the environment. The superior flexibility also enables deposition of 2D films on scanning probe centilevers, creating a conceptually new sensing methodology based on a continuous tuning of the twist angle between different 2D materials. Commensurate states forming at specific alignment angles lead to a significant reconstruction of the electronic band structures, creating resonant conditions for electron tunnelling.

\begin{figure*}[]
\includegraphics[width=18cm]{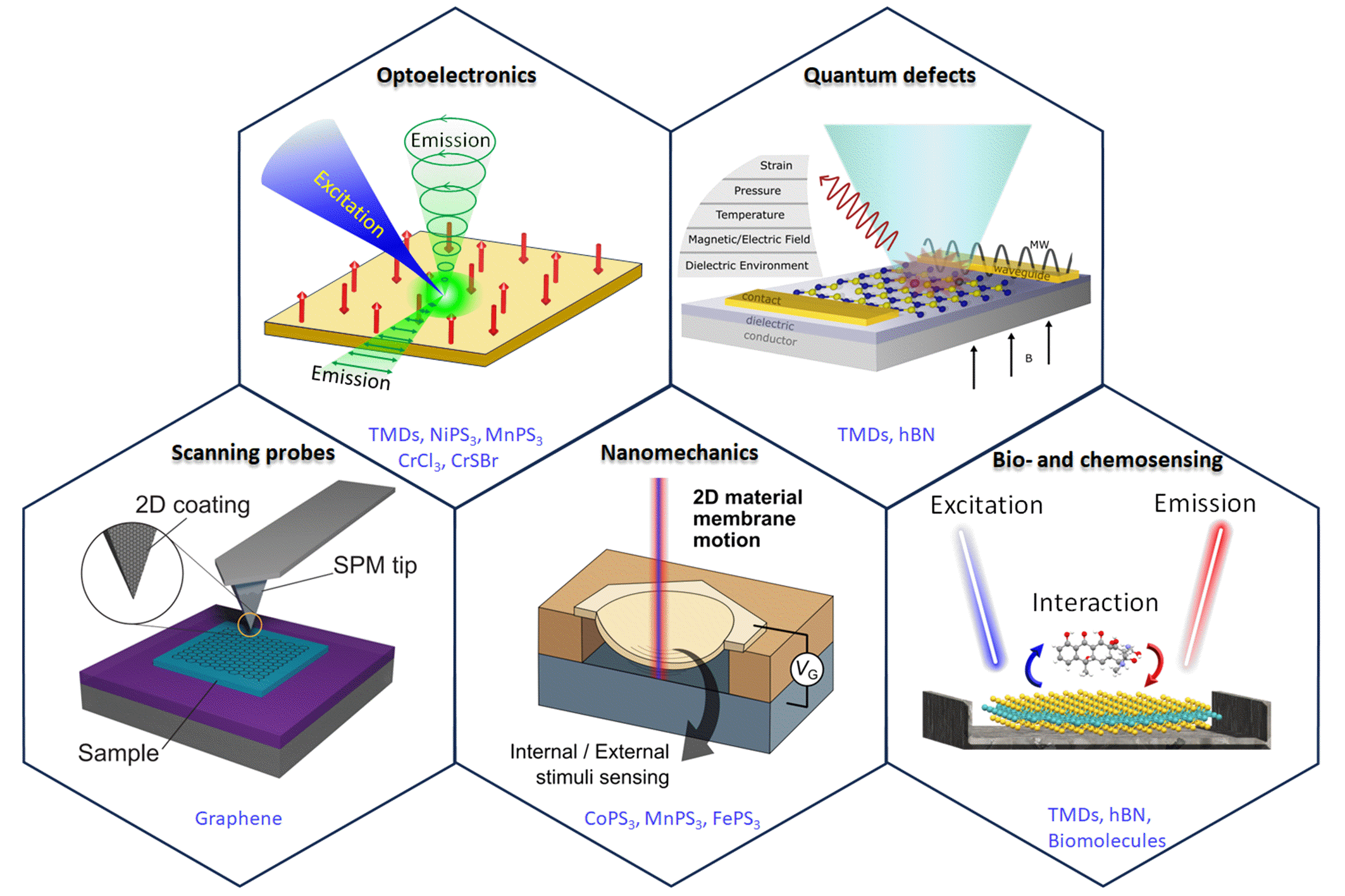}
\caption{ \textbf{Multiproperty sensing in two-dimensional (2D) materials.} The schematic representation of various sensing modalities where 2D materials provide unique advantages. These include: 1) intertwined optoelectronic and magnetic properties, 2) quantum characteristics of defect centers at the atomic scale for ultimate minaturization of environmental sensing, 3) integration of 2D materials with scanning probes for tailored tip-surface interactions, 4) nanomechanical motion of 2D nanodrums coupled to multiple quantum degrees of freedom, and 5) bio- and chemosensing at a large surface-to-volume ratio.}
\label{fig:fig0_content}
\end{figure*}

Here, we review the progress in 2D sensing technology, focusing on five areas where 2D materials provide apparent benefits. Section~\textbf{I.} focuses on sensing based on optoelectronic methods realized in individual materials, heterostructures, and devices. Section~\textbf{II.} presents atomic-scale sensing with single photon emitters. Section~\textbf{III.} explores recent advancements in 2D coating for scanning probe cantilevers, which create novel sensing methodologies such as the quantum twisting microscope. Section~\textbf{IV.} summarizes the achievements in detecting phase transitions and external stimuli with 2D nanodrums. Section~\textbf{V.} demonstrates the applications of 2D materials for bio- and chemosensing. A schematic representation of distinct sensing modalities enabled and/or enhanced by 2D materials is depicted in \textbf{Fig.~\ref{fig:fig0_content}}.

\section{Intra– and intermaterial sensing by optoelectronic means}

\begin{figure*}[]
\includegraphics[width=17cm]{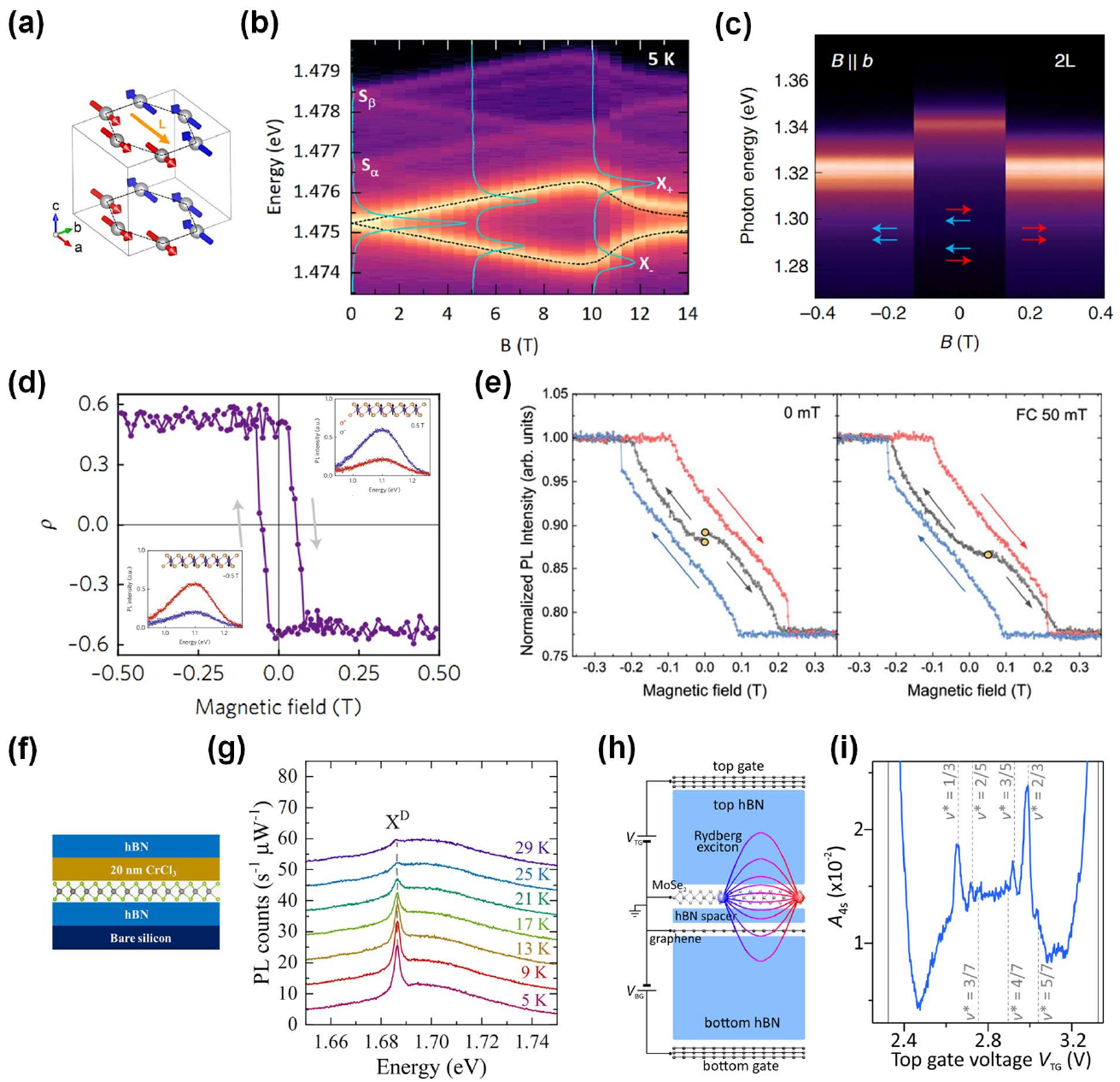}
\caption{ \textbf{Sensing by optoelectronic effects in 2D materials.} \textbf{(a)} The schematic magnetic structure of NiPS$_3$ in the antiferromagnetic phase as a representative structure of the van der Waals magnetic materials. Grey spheres with associated red and blue arrows represent Ni$^{2+}$ ions and the spin direction of two sublattices, respectively. \textbf{(b)} False color map of low temperature photoluminescence (PL) of NiPS$_3$ film as a function of the magnetic field applied along the easy axis. A few representative PL spectra at different magnetic fields are also shown. \textbf{(c)} False color map of low temperature PL of CrSBr bilayer as a function of the magnetic field applied along the easy axis. \textbf{(d)} Circular polarization ($\rho$) resolved PL intensity hysteresis loop as a function of magnetic field applied along the crystal c-axis of CrI$_3$. The inset shows the circular polarization resolved PL spectra at -0.5~T and +0.5~T at 15~K. \textbf{(e)} PL intensity hysteresis loops revealing exciton–skyrmion coupling in 170~nm thick CrBr$_3$ at zero-field cooling and 50~mT field cooling. \textbf{(f)} The schematic side view of the hBN encapsulated CrCl$_3$/WSe$_2$ heterostructure. \textbf{(g)} The temperature-dependent PL spectra of the CrCl$_3$/ML-WSe$_2$ heterostructure demonstrating dark exciton resonance. \textbf{(h)} Schematic illustration of the MoSe$_2$/graphene heterostructure for sensing the compressibility of graphene electrons with Rydberg excitons. \textbf{(i)} The top-gate voltage (V$_{TG}$) dependent amplitude of the 4s exciton exhibiting several prominent maxima, revealing a multitude of fractional quantum Hall states with denominators of 3, 5, and 7 (as indicated by the vertical dashed lines). Panels \textbf{a} and \textbf{b} are adopted from Ref.~[25], American Physical Society; panels \textbf{c} and \textbf{d} are adopted from Ref.~[26, 28], Springer Nature Limited; panel \textbf{e} is adopted from Ref.~[30], Wiley Online Library; panels \textbf{f} and \textbf{g} are adopted from Ref.~[34], Royal Society of Chemistry; panels \textbf{h} and \textbf{i} are adopted from Ref.~[40], American Chemical Society. }
\label{fig:fig1_optics}
\end{figure*}

2D semiconductors host excitons that can serve as effective sensors for quantum phases due to their strong light-matter interactions and sensitivity to external perturbations \cite{wang2018colloquium, arora2015, Arora_2015_2, Smolenski2016, Slobodeniuk_2019, Koperski_2019, Arora_2019, Zultak2020, 2DPhotonics_Roadmap}. The unique properties of excitons, including their binding energies and spatial characteristics, make them valuable for exploring quantum phenomena, such as quantum phase transitions, spin physics, and the formation of nontrivial magnetization textures. Van der Waals magnetic materials have attracted significant interest due to their ability to retain intrinsic magnetic order down to the monolayer limit \cite{gibertini2019magnetic, Wang2022, huang2017layer}. These materials exhibit a broad variety of magnetic ground states, stabilized by the interplay of exchange interactions and crystalline anisotropy. Application of external stimuli, such as temperature, magnetic field, and pressure, can modulate the spin orientation, inducing phase transitions between distinct magnetic states. Many of these magnetic materials have strong excitonic responses in addition to magnetism \cite{acharya2022real, acharya2023theory}. The coupling between intrinsic optoelectronic and magnetic properties makes them promising platforms for developing non-contact optical magnetic sensing techniques.

As recently reported, 2D antiferromagnetic materials exhibit a sharp optical transition when cooled below the N\'{e}el temperature \cite{Gnatchenko2011, Kang2020, son2022multiferroic}. Although the origin is controversial, the transition is confirmed to be related to a spin flip exciton that serves as a sensitive probe of magnetic ordering \cite{He2024}. The particular exciton often exhibits characteristic polarization that depends on both the spin orientation and the detection geometry. For instance, in NiPS$_3$, the emission is found to be highly linearly polarized, with the electric field vector perpendicular to the spin direction  \cite{Wang2021}. Consequently, any spin reorientation driven by the magnetic phase transition is observed as a rotation of the exciton's polarization axis. This spin reorientation, induced by the magnetic field, also appears as an anomalous splitting of the exciton as shown in \textbf{Fig.~\ref{fig:fig1_optics}(a,b)} \cite{jana2023magnon}. The energy of the exciton as a function of external magnetic field enables the prediction of the critical field for different magnetic phase transitions of the material. 

The spin-allowed localized excitons also exhibit clear signatures of the underlying magnetic order, making them highly sensitive probes of magnetic states. Changes in magnetic ordering directly influence the bandgap and exciton binding energy, highlighting a strong magneto-electronic coupling. This phenomenon has been well demonstrated in the A-type antiferromagnet CrSBr through the study of Wannier-Mott excitons \cite{wilson2021interlayer}. In this material, spins in each layer are ferromagnetically coupled, while the adjacent layers are coupled via a weak antiferromagnetic exchange interaction. A small magnetic field is found to compete with the anisotropy and antiferromagnetic exchange parameters, transforming them to the ferromagnetic states, which can be sensed by the energy shift of the excitons in the infrared range, as shown in \textbf{Fig.~\ref{fig:fig1_optics}(c)}. Additionally, the interlayer antiferromagnetic exchange interaction is tuned by applying hydrostatic pressure, which alters the magnetic phase transition points concurrently with revealing new magnetic phases, manifesting as measurable changes in exciton energy \cite{pawbake2023magneto}.

Spin-allowed optical transitions arising from localized, parity-forbidden d–d orbitals centered on metal atoms are also sensitive to magnetic ordering, making them valuable for optical sensing applications. In two-dimensional magnetic materials such as CrI$_3$ and CrBr$_3$, these transitions are found to be circularly polarized, particularly when excited resonantly, with the helicity determined by the underlying magnetic state \cite{seyler2018ligand, grzeszczyk2023strongly}. CrI$_3$, for instance, exhibits ferromagnetic ordering within each monolayer but antiferromagnetic interlayer coupling in the bilayer configuration. As a result, monolayers display a pronounced degree of circular polarization, which diminishes in the bilayer. Further, the degree of circular polarization in a monolayer vanishes above the Néel temperature when spin order is lost. The application of an external magnetic field demonstrates a hysteresis indicative of the intrinsic ferromagnetic ordering as shown in \textbf{Fig.~\ref{fig:fig1_optics}(d)}. CrBr$_3$ is an insulating 2D ferromagnet capable of hosting diverse spin textures, including stripe domains and skyrmion lattices. Magneto-photoluminescence spectroscopy resolved by circular polarization enables direct optical sensing of local magnetization states in CrBr$_3$ (\textbf{Fig.~\ref{fig:fig1_optics}(e)}) \cite{grebenchuk2024topological, Grebenchuk_2024, grebenchuk2025correlations}. Under different magnetic field histories (zero-field cooling and field cooling), the photoluminescence intensity exhibits distinct hysteresis behaviors corresponding to stripe or skyrmion spin textures. This intramaterial sensing mechanism arises from exciton–spin coupling: the spin alignment modulates optical selection rules, allowing the PL spectrum to optically encode the underlying magnetic state. This technique provides a basis for imaging and identifying topologically nontrivial spin states using light alone, without electrical contacts. It paves the way for optical magnetometers or skyrmion-based memory readout schemes that rely purely on photonic detection.

Being 2D in nature, the heterostructures composed of magnetic material and semiconducting transition-metal dichalcogenides (TMDs) offer a versatile platform for sensing magnetic proximity effects through optical means. Various excitonic features that dominate the optical response of monolayer TMDs exhibit valley splitting, polarization selectivity, and the activation of dark exciton states even in the absence of external stimuli, serving as a fingerprint for the proximity-induced magnetic field in these heterostructures. For instance, in bare monolayer WSe$_2$, the exciton photoluminescence intensity from K$^+$ and K$^-$ valleys is identical. However, when interacting with a ferromagnetic CrI$_3$ layer, a time-reversal symmetry is broken. This leads to an imbalance in the photoluminescence intensity, directly correlating with the magnetization of CrI$_3$ \cite{zhong2020layer}. The degeneracy of the exciton energy from K$^+$ and K$^-$ valleys is lifted due to the effective exchange field from the magnetic material. An effective magnetic exchange field of approximately 20~T was reported in such systems, inferred from the observed valley polarization and Zeeman splitting of excitons without any externally applied magnetic field \cite{seyler2018ligand}. The magnetic phase transitions and hysteresis inherent to the ferromagnetic layer are also directly reflected in the valley-selective exciton properties of the adjacent TMD layer.  Furthermore, ferromagnetic CrCl$_3$ with an in-plane easy spin axis is demonstrated to induce brightening of dark excitons in monolayer WSe$_2$ through the intrinsic in-plane component of the proximity magnetic field, as shown in \textbf{Fig.~\ref{fig:fig1_optics}(f,g)} \cite{kipczak2025interplay}. 

In a typical p-n junction device configuration, WSe$_2$-CrI$_3$ hybrid has recently been demonstrated to act as a self-powered light helicity detector \cite{chen2024spin}. The enhanced detection capability is attributed to the synergistic interaction between the spin-filtering properties of CrI$_3$, the valley-selective electronic transitions in WSe$_2$, and spin-dependent charge transfer processes at the interface. This interplay facilitates selective excitation and separation of spin-polarized carriers, leading to a measurable photocurrent that reflects the helicity of the incident light. In contrast to electrons in graphene, TMD heterostructures (especially twisted TMD structures \cite{Alexeev2019}) offer an optically active system for exploring the correlated many-body states, such as Mott insulator or Wigner crystal states \cite{tang2020simulation}. Due to the large dipole moment of intra- and interlayer excitons, the optical detection of such correlated states is possible with enhanced sensitivity. Using WSe$_2$ as a proximity sensor, multiple correlated insulating states were optically explored in the WSe$_2$-WS$_2$ moir\'{e} superlattice by detecting the resonance energy and oscillator strength of excited exciton states in the sensor \cite{xu2020correlated}. By altering the effective dielectric constant or by enhancing exciton-carrier many-body interactions, excitons interact with associated electronic states, enabling optical spectroscopy to probe charge correlations. Such an approach has the benefit of studying spin and charge dynamics, magnetic order \cite{tang2020simulation}, and other correlated physics that are not directly accessible by transport measurements due to the presence of valley-dependent and valley-independent carrier-exciton interactions.

Optical spectroscopy can also detect certain exciting electrical transport processes, such as the fractional quantum Hall (FQH) effect of graphene \cite{polshyn2018quantitative}. In particular, Rydberg excitons in TMD semiconductors, characterized by a series of excited states (2s, 3s, and higher), possess Bohr radii significantly larger than the thickness of the monolayer, making them highly sensitive to the surrounding dielectric environment. When compared to traditional transport techniques, Rydberg exciton spectroscopy provides great spatial resolution and reduces the influence of spatial inhomogeneities inherent to most van der Waals heterostructures. Rydberg excitons of the monolayer of MoSe$_2$ have been demonstrated to detect graphene FQH states even without direct electrical contact (\textbf{Fig.~\ref{fig:fig1_optics}(h,i)}) \cite{popert2022optical}. These excitons are highly sensitive to the dielectric environment of neighboring graphene layers, acting as quantum proximity sensors.

\section{Sensing with quantum defects in 2D materials}
\label{section:defects}

\begin{figure*}[]
    \centering
    \includegraphics[width=14 cm]{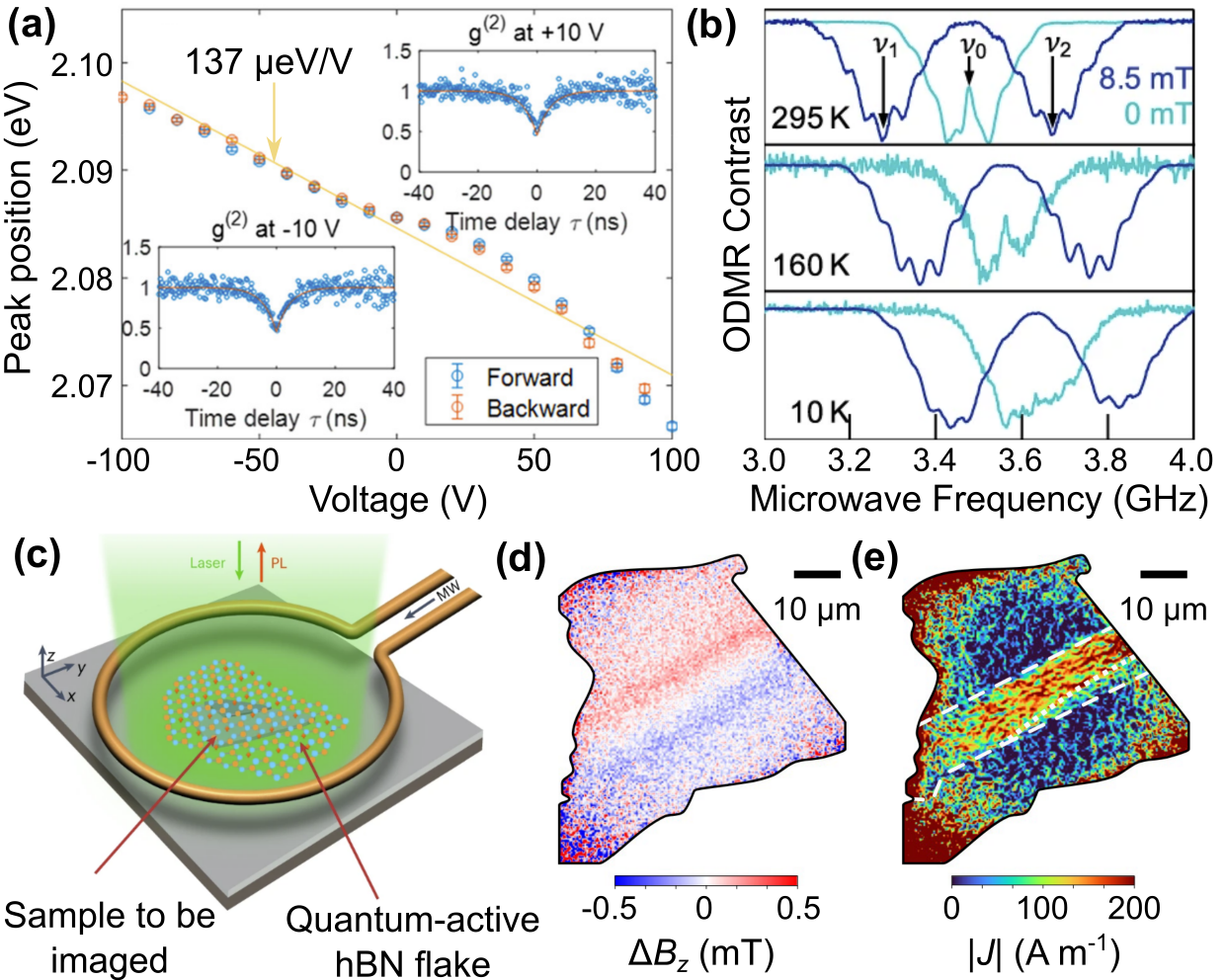}
    \caption{\textbf{Sensing with quantum defects.}
    \textbf{(a)} Observation of the room-temperature giant Stark effect in an hBN SPE. The blue and orange dots correspond to experimental data obtained during the forward and backward sweeping of voltages, respectively.  A tuning efficiency of 137~$\upmu$eV/V is obtained by linear regression (yellow solid line). Insets show second-order correlation functions $g^{(2)}$ measured at $\pm$10~V, confirming that the single photon emission persists under external electric fields.
    \textbf{(b)} Temperature sensing via optically detected magnetic resonance (ODMR), illustrated through the temperature dependence of the zero-field splitting. ODMR spectra are shown with (dark blue) and without (cyan) an external magnetic field, measured at 295, 160, and 10~K.
    \textbf{(c)} Schematic of an hBN flake containing \vb{} spin defects (small red arrows) placed on a sample under investigation. The defects are excited by laser and microwave (MW) fields, and their photoluminescence (PL) is imaged with a camera for spatially resolved ODMR.
    \textbf{(d)} Experimental microscopic ODMR map of the stray magnetic field component ($\Delta B_z$) under $I=3$~mA current measured according to the schematic in \textbf{c}. The sample is a few-layer graphene ribbon covered with a 70~nm-thick hBN flake and contacted with gold electrodes.
    \textbf{(e)} Current density map (normalized $|J|$) reconstructed from \textbf{d}. Panel \textbf{a} is adapted from Ref. [68], American Chemical Society; panel \textbf{b} is adapted from Ref. [66],  Springer Nature Limited; panels \textbf{c}, \textbf{d}, and \textbf{e} are adapted from Ref. [80], Springer Nature Limited.}
    \label{fig:defects}
\end{figure*}

Quantum sensing using quantum emitters \cite{Koperski2015, Srivastava2015, He2015, Chakraborty2015, Tran2016, KOPERSKI_2018, Koperski2021,  LITVINOV_2025} in 2D materials offers a compelling route toward ultra-high spatial resolution. Individual quantum emitters embedded in monolayer systems \cite{Loh2024} can be positioned in direct proximity to surfaces and interfaces, making them ideal for probing nanoscale fields, strain, and dielectric variations. The atomic-scale thickness, strong light–matter interactions, and external field tunability of 2D materials position them as uniquely suited for integrated quantum sensing applications. Among the growing list of 2D candidates, TMDs and hexagonal boron nitride (hBN) have emerged as complementary platforms with distinct strengths.

TMD monolayers such as WSe$_2$ and WS$_2$ host localized defect states that function as single-photon emitters (SPEs), often activated by deterministic strain engineering. Local deformation introduces potential wells that modify the band structure and localize excitons at defect sites, enabling bright and stable quantum emission~\cite{iff2019strain,abramov2023photoluminescence,yu2025dynamic,rosenberger2019quantum}. These defect-bound excitons are extremely sensitive to the local potential landscape, and their spectral shifts under strain offer a quantitative readout, highlighting their potential as nanoscale strain sensors. Beyond strain, these defect states are also responsive to electric fields. Individual defects in the monolayer WSe$_2$, embedded within van der Waals heterostructures that incorporate hBN and graphene \cite{Grzeszczyk2024, Howarth2024, Loh2024_2}, show narrow photoluminescence and electroluminescence peaks that shift under applied vertical bias. The observed Stark shifts (0.4~meV/V) allow extraction of permanent dipole moments, confirming sensitivity to the electrostatic environment~\cite{schwarz2016electrically}. Similarly, electrically driven emission in WS$_2$ exhibits spectral modulation tied to defect charging and gate tuning~\cite{schuler2020electrically}. In addition to localized defect states, moir\'{e} excitons in TMD heterobilayers formed by stacking two monolayers with a slight twist angle or lattice mismatch represent another class of quantum-confined states highly sensitive to external fields. Due to their spatially modulated potential landscape, these excitons exhibit tunable optical transitions and exhibit large Stark shifts (from their interlayer dipolar nature) up to 429~meVnmV$^{-1}$\cite{baek2020highly}. SPEs in TMDs also display significant sensitivity to dielectric surroundings. In monolayer WSe$_2$, narrow-linewidth, linearly polarized, and spatially localized SPEs exhibit strong spectral shifts depending on the substrate (e.g., SiO$_2$ vs. sapphire) or encapsulation, due to variations in local dielectric screening~\cite{dang2020identifying}. These changes modulate exciton binding energies and charge localization potentials. While initial demonstrations have primarily focused on the controllability and tunability of quantum emitters in TMDs, their pronounced spectral sensitivity to external fields and dielectric environments highlights their emerging potential as nanoscale sensors

In contrast to TMDs, hBN has become a central platform for quantum sensing, largely due to the diverse range of optically active quantum emitters. Crucially, hBN hosts both spin-active and optically bright defects, including color centers such as carbon-related emitters \cite{Koperski2020, Huang2022, WU_2024} and negatively charged boron vacancy (\vb{})\cite{durand2023optically, gottscholl2021room,gottscholl2021spin}. Compared to traditional systems like nitrogen-vacancy (NV) centers in diamond or quantum dots, hBN emitters offer stronger Stark responses (reaching 54~meVnmV$^{-1}$ at low temperature~\cite{noh2018stark} and 43~meVnmV$^{-1}$ at ambient conditions~\cite{xia2019room} (see \textbf{Fig.~\ref{fig:defects}(a)}) and simpler operation without temperature or microwave control~\cite{xia2019room,gruber1997scanning}. While NV [or negatively charged NV (\nv{})] centers are spin active and widely used in magnetometry, their sensitivity to electric fields is limited due to their high symmetry and electronic structure, leading to a requirement of microwave excitation~\cite{dolde2011electric,barson2017nanomechanical}. Quantum dots also exhibit tunable optical transitions but are prone to instability and spectral diffusion at room temperature~\cite{kuhlmann2015transform}. In contrast, quantum emitters in hBN show distinct responsiveness to both electric fields and dielectric environments. For example, carbon-related defects demonstrate strong zero-phonon line shifts, modified phonon sidebands, and linewidth broadening when their local dielectric surroundings (such as the number of hBN layers or the substrate permittivity) are altered~\cite{badrtdinov2023dielectric}. These changes result from modulation of intra-defect Coulomb interactions, allowing such defects to serve as highly localized dielectric sensors. Additionally, integrating hBN SPEs into optical cavities or waveguides enables spectral engineering through refractive index contrast, offering enhanced coupling efficiency and stability, critical factors for photonic integration and environmental sensing~\cite{li2021integration}.

Multiple spin defects have been studied in hBN, including structures with a triplet ground state (spin-1 defects)\cite{gottscholl2020initialization, stern2024quantum,stern2022room}. The most explored example is a \vb{} complex which has a zero-field splitting (D) of around 3.47~GHz with out-of-plane anisotropy. An existence of well-established technology of deterministic positioning of boron vacancies\cite{liang2023high} makes this defect an attractive platform for multiple sensing applications. The reliable creation and readout of \vb{} centers have already enabled applications in the domain of sensing temperature and pressure variations~\cite{gottscholl2021spin} as presented in \textbf{Fig~\ref{fig:defects}(b)}. \vb{} centers have also been used to detect ferromagnetic resonance in YIG~\cite{das2024quantum} and image ferromagnetic domains in 2D magnets\cite{healey2023quantum,huang2022wide} as schematically depicted in \textbf{Fig.~\ref{fig:defects}(c)}, with example mapping of the ODMR response and current shown in \textbf{Fig.~\ref{fig:defects}(d,e)}. Moreover, the transverse splitting (E) has been found to arise mainly due to the formation of an electric dipole formed by a negatively charged vacancy and positively charged environment, giving \vb{} centers sensitivity to nearby charges within 10~nm~\cite{durand2023optically}. While the magnetic sensitivity of the \vb{} centers is estimated to be around 85~$\upmu T/\sqrt{Hz}$, other recently identified defects, such as carbon-related SPEs\cite{stern2024quantum}, exhibit improvements of almost two orders of magnitude. Although these values are still three orders of magnitude below state-of-the-art \nv{} centers in isotopically purified diamond, the three-dimensional nature of diamond limits the proximity of \nv{} centers to target materials, reducing their depth below 5~nm (without spin coherence time being affected). In contrast, spin-active defects in hBN can be integrated directly with other 2D materials, enabling sensing at true surface proximity. ODMR signals have been successfully observed in few-layer hBN\cite{durand2023optically}, supporting this approach. Moreover, as the magnetic sensitivity of \vb{} is mainly limited by spin coherence time affected by hyperfine coupling, it can be improved by using decoupling protocols\cite{rizzato2023extending,ramsay2023coherence} or by using different nuclear isotopes of hBN host lattice\cite{gong2024isotope}. 
Furthermore, carbon-related defects in hBN exhibit multiple anisotropy axes, enabling vector magnetometry~\cite{m2025single}.

Although most studies to date have focused on the controllability and tunability of quantum emitters, the same physical principles, i.e., extreme sensitivity to local fields, strain, and dielectric environments, make them powerful tools for nanoscale sensing. TMDs, with their strong field responsiveness and compatibility with electrical platforms, are well-suited for strain, electric field, and dielectric sensing. In contrast, hBN hosts stable, room-temperature spin-active defects that function as truly multipurpose sensors capable of detecting magnetic fields, temperature, pressure, and local charge variations at the nanoscale.

\section{Integration of 2D materials onto scanning probes for sensing}

\begin{figure*}[]
    \centering
    \includegraphics[width=1\linewidth]{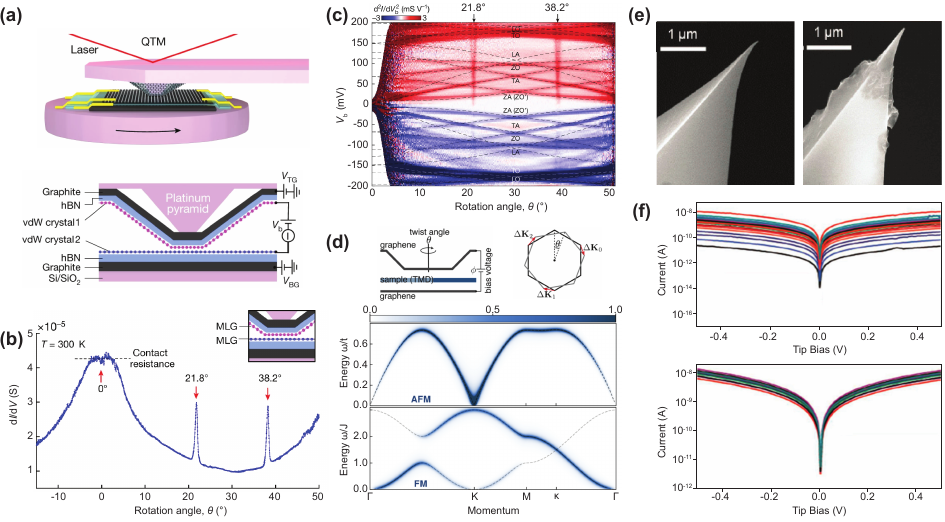}
    \caption{\textbf{2D materials for scanning probe microscopy.} 
    \textbf{(a)} A schematic illustration of a quantum twisting microscope (QTM). It consists of an AFM cantilever equipped with a plateaued pyramid tip covered with a 2D material and a sample (e.g., a 2D film deposited on a Si/SiO$_2$ substrate) mounted on a rotational stage.
    \textbf{(b)} Differential conductance $dI/dV$ measured in a QTM as a function of an angle $\theta$ between two monolayer graphene films in direct contact.
    \textbf{(c)} The second-order differential conductance $d^2I/dV^2$ measured as a function of bias voltage between two graphene films and the relative angle between them. It reveals resonances attributed to acoustic (TA, ZA) and optical (LO, TO, ZO) phonon modes.
    \textbf{(d)} - Theoretical proposal for studying magnetism using QTM. It is based on an intermediate layer sandwiched between two graphene films. The biasing and angular control knobs enable exploration of a wide range of unique conditions modulating spin-spin interactions. The theory predicts a dynamical spin-structure factor related to the antiferromagnetic (AFM) and ferromagnetic (FM) phases.
    \textbf{(e)} Scanning electron microscopy (SEM) images of a commercial Pt-varnished cantilever (left) and a cantilever with graphene coating (right). 
    \textbf{(f)} $IV$-curves for 18 different Au cantilevers (top graph) and for 18 Au cantilevers with graphene coating (bottom graph). $IV$-curves were obtained for Au/octanemonothiol/Au and Au/octanemonothiol/graphene/Au junctions, respectively. Panels \textbf{a} and \textbf{b} are adapted from Ref.~[92], Springer Nature Limited; panel \textbf{c} is adapted from Ref.~[94], Springer Nature Limited; panel \textbf{d} is adapted from Ref.~[97], American Physical Society; panel \textbf{e} is adapted from Ref.~[86], Royal Society of Chemistry; panel \textbf{f} is adapted from Ref.~[87], John Wiley $\&$ Sons, Inc.
    } 
    

    \label{fig:spm}
\end{figure*}

2D materials can be deposited onto or integrated with scanning probes, such as atomic force microscopy (AFM) cantilevers. Currently, several methods have been realized to place 2D materials, mostly graphene, on AFM probes~\cite{hui2017graphene}. These include chemical vapor deposition (CVD) on tip~\cite{wen2012multilayer}, direct graphene-like thin film sputtering deposition followed by \textit{in situ} annealing ~\cite{pacios2016direct}, transfer of pre-grown CVD graphene~\cite{lanza2012graphene}, mold-assisted transfer of CVD-grown graphene~\cite{martin2013graphene}, coating with liquid-phase graphene flakes~\cite{hui2016moving}, and polymer membrane transfer techniques on flat pyramid-shaped tips~\cite{inbar2023quantum}. In comparison to existing cantilever coatings, coating AFM probes with 2D materials can allow one to achieve improvements to existing measurement techniques and promote the development of entirely new sensing methods.

An innovative example is the realization of the quantum twisting microscope (QTM)~\cite{inbar2023quantum, ke2024proposed, birkbeck2025quantum}. In QTM, one layer of a 2D material is placed on the AFM cantilever with a flat pyramid-shaped tip, while another layer is placed on a substrate, which is mounted on the holder with rotation capability. This enables \textit{in situ} tuning of twist angle, as schematically illustrated in \textbf{Fig.~\ref{fig:spm}(a)}. This technique allows probing and studying electronic states in moir\'e structures as a function of a twist angle. Moir\'e physics enables the study of a wide range of phenomena, often involving correlated electronic states. These include superconductivity in magic-angle twisted bilayer graphene, Mott insulating behavior, and emergence of ferroelectric domains in hBN and TMDs~\cite{he_moire_2021}. However, device fabrication constitutes a bottleneck for a systematic exploration of a large parameter space, as typical samples exhibit a singular twist angle. QTM addresses this challenge, allowing measurement of the twist angle dependence on a single sample. QTM has already revealed twist-angle-dependent electronic properties in twisted graphene, featuring peaks at 21.8$^\circ$ and 38.2$^\circ$ twist angles which correspond to the formation of commensurate stacking of two layers both in real and momentum spaces (\textbf{Fig.~\ref{fig:spm}(b)}). This leads to quantum coherence between the layers, promoting higher tunneling from one layer to the other. 

Additionally, the QTM-based method was developed to directly measure phonon dispersion and electron–phonon coupling through inelastic momentum-resolved spectroscopy \cite{birkbeck2025quantum, xiao2024theory}, as shown in \textbf{Fig.~\ref{fig:spm}(c)}. This method also allowed for the detection of the layer-antisymmetric low-energy acoustic phason mode, which arises from sliding or shear between the layers of moir\'e lattice. The study reveals that the phason mode is strongly coupled to the electrons tunneling between the layers, revealing its contribution to the electronic behavior of twisted bilayer graphene.

Several theoretical studies further propose employing QTM for sensing a wide range of electronic, magnetic, and structural phenomena~\cite{xiao2024theory, pichler2024probing, wei2025dirac}. QTM was proposed as a method for sensing magnetism in moir\'e heterostructures and distinguishing between ferromagnetic and antiferromagnetic ordering by analyzing single-particle and collective-excitation spectra~\cite{pichler2024probing} (\textbf{Fig.~\ref{fig:spm}(d)}). By placing a TMDC monolayer between two graphenes and creating a particular twist angle between them, it is possible to stabilize a generalized Wigner crystal state formed in the moiré superlattices of the TMDC heterostructure. In such a system, magnetic ordering can arise from correlated behavior of electrons in Wigner crystals, and it is determined by the competition of superexchange and direct exchange interactions.

2D materials can also be used for improving existing techniques. For example, graphene-coated AFM cantilevers (\textbf{Fig.~\ref{fig:spm}(e)}) have demonstrated excellent electrical performance, enabling more accurate current measurements and improved lateral resolution in conductive AFM. Additionally, the hydrophobic and chemically inert nature of graphene reduces tip wear and enhances measurement consistency~\cite{shim2012multifunctional, hui2017graphene} (\textbf{Fig.~\ref{fig:spm}(f)}).

In addition to graphene, hexagonal boron nitride (hBN) can also be implemented on the cantilevers, providing sensing functionalities similar to nitrogen vacancy (NV) magnetometry based on diamond tips \cite{taylor2008high, maletinsky2012robust}. Point defects in hBN layers are sensitive towards magnetic fields and temperature gradients~\cite{healey2023quantum}. This can allow for performing magnetic sensing at the nanoscale and visualizing magnetic structures, such as magnetic domains, skyrmions, or stray fields generated by current flow.
Additional advantages of hBN-defect-based technique are higher sensitivity to electric fields and dielectric environment, which can give even broader possibilities in application for sensing compared to diamond NV-centers, as was discussed in more detail in the section ~\ref{section:defects}.

\section{2D membranes as sensors}

\begin{figure*}[]
    \centering
    \includegraphics[width=\linewidth]{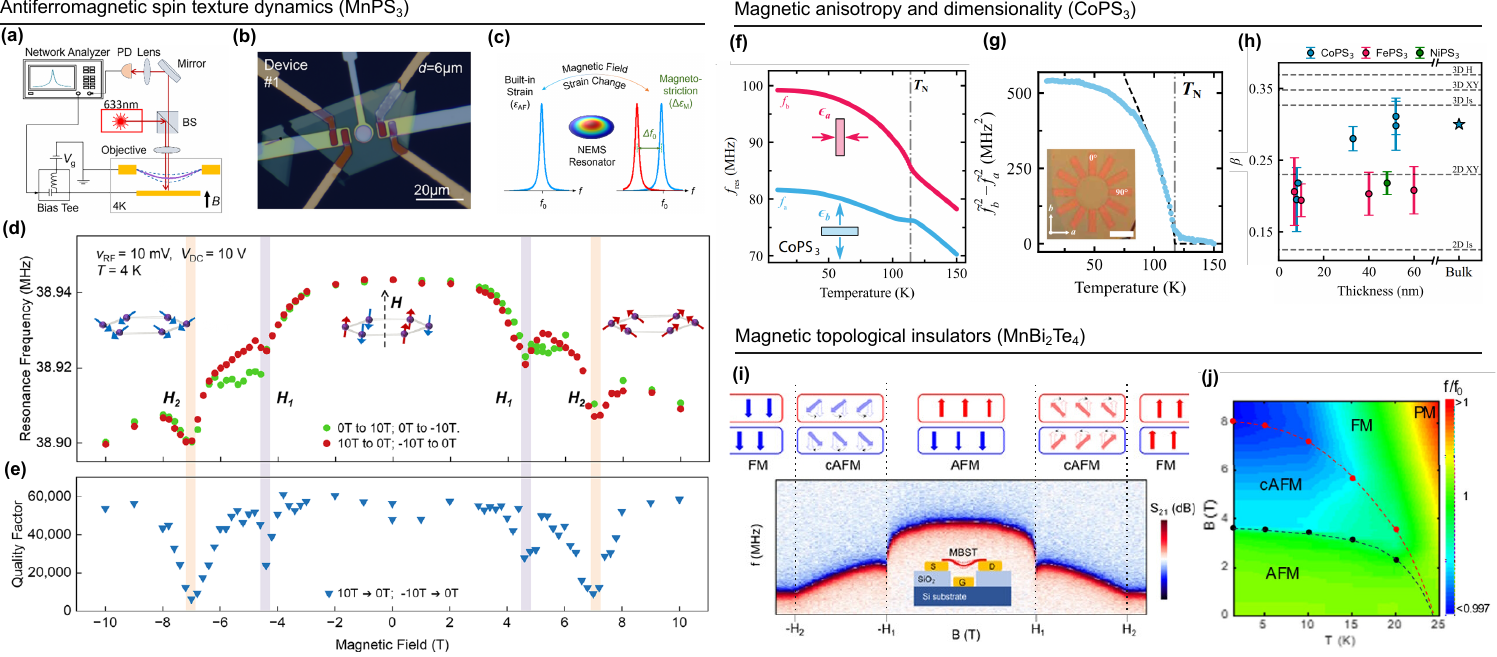}
    \caption{\textbf{Sensing magnetism with nanomechanical membrane resonators.} \textbf{(a)} Optical interferometry system to measure the driven response of the resonators. The applied magnetic field is along the out-of-plane direction of the suspended membrane. \textbf{(b)} The microscopy image of a MnPS$_3$ nanoelectromechanical resonator. Scale bar: 20 $\upmu$m. \textbf{(c)} Conceptual illustration of magnetostriction-induced resonance frequency shift due to the spin-flop transition in MnPS$_3$ mechanical resonators. \textbf{(d)} Evolution of resonance frequency as a function of out-of-plane magnetic field measured between $\pm$10 T. \textbf{(e)} Measured Q factor versus magnetic field shows strong dissipation at the sharp transitions. \textbf{(f)} Temperature dependence of a CoPS$_3$ rectangular membrane. The arrows show the dominant magnetostrictive strain contributions for the corresponding cavities. \textbf{(g)} Difference of the frequency squared proportional to the antiferromagnetic order parameter $\mathrm{L^2}$. The dashed black line is a power-law fit through the data close to T$\mathrm{_N}$. Inset: Optical image of the rectangular membranes array for a CoPS$_3$ sample. Scale bar: 12 $\upmu$m. \textbf{(h)} Average critical exponent, $\beta$, and T$\mathrm{_N}$, of MPS$_3$ resonators plotted as a function of thickness. Temperature dependence of magnetic phase transition in MBST probed by resonance frequency. Color plots of RF transmission signal S$_{21}$ as functions of driving frequency and magnetic field at a fixed gate voltage. \textbf{(j)} Magnetic phase diagram derived from the $f\mathrm{(B, T)}$. Color plot of $f(\mathrm{B})/f(\mathrm{0})$ as a function of magnetic field and temperature. Panels \textbf{a-e} are adapted from Ref.~[119], Springer Nature Limited; panels \textbf{f-h} are adapted from Ref.~[118], Springer Nature Limited; panels \textbf{i} and \textbf{j} are adapted from Ref.~[120], American Chemical Society.}
    \label{Fig:membranes}

    
\end{figure*}

2D materials have a remarkable ability to sense and respond to the smallest physical stimuli. Even being just a few atomic layers thick, membranes made of these materials combine extreme mechanical responsivity to forces with rich and tunable magnetic and electronic behaviour. When these 2D membranes are suspended over a cavity, they act as nanoscale resonators, whose resonance frequency is governed mainly by the tension accumulated in the layer. In this way, any subtle force acting on it can leave a measurable fingerprint on its motion. In recent years, researchers have transformed these ultrathin systems into powerful probes by tracking changes in resonance frequency, motion amplitude, or mechanical energy dissipation. This approach allows for probing both external stimuli and inner material physics, advancing the frontiers of precision sensing, ranging from detecting biological vitality to studying condensed matter phase transitions that are challenging to access otherwise.

One of the most compelling demonstrations of this capability to detect external forces is in the detection of biological motion. Suspended graphene membranes, configured as drums, have been shown to detect nanometric vibrations generated by individual bacteria \cite{Irek2022}. These subtle perturbations manifest as amplitude noise in the membrane's mechanical resonance. Upon introduction of antibiotics, bacterial activity ceases, and the resulting spectral quiescence offers a direct indicator of microbial viability,  providing a label-free approach to single-cell antimicrobial susceptibility testing. Similarly, 2D MXenes have been recently made into suspended Ti$_3$C$_2$T$_x$ membrane resonators~\cite{Tan2022monolayer, Xu2022electrically, Wan2024elastic}, utilizing their electrical tuning, intrinsic piezoelectric and strong surface adsorption properties, and enabling their operation as high-sensitivity mass detectors with sub-zeptogram resolution \cite{Tan2022monolayer}. Photothermal sensing constitutes another important modality. When photons are absorbed by a 2D membrane, the resulting local heating induces a measurable shift in resonance frequency \cite{femtowatKirchhof2023,spectroscKirchhof2022,Aguila2022}. These devices leverage the interplay between thermal conductivity and high thermal expansion coefficients of suspended 2D materials to achieve outstanding optical sensitivity. Graphene and transition metal dichalcogenide resonators have thus been utilised as bolometers and optical detectors \cite{Blaikie2019, bolometerChen2022} as well as probes for exciton-optomechanical coupling \cite{Xie2021}, and nanomechanical spectroscopy \cite{spectroscKirchhof2022} reaching sensitivity down to femto-Watts \cite{femtowatKirchhof2023}.

Beyond interactions with the external world, 2D membranes serve as sensitive probes of their own intrinsic material-related degrees of freedom. A prominent application established in recent years is in tracking condensed matter phase transitions, particularly magnetic \cite{Siskins2020,FaiMakJiang2020,CGTSiskins2021,Zhang2022,LpezCabrelles2021,Li2022, Maurits2023, Yousuf2025, Liu2025, Cheng2024,Fei2024, Li2023, Houmes2023, NonlinSiskins2025}, electronic \cite{Lee2021,Siskins2020,Sengupta2010,Will2017,Sahu2022}, or structural \cite{Mao2024,Chaste2020}, via shifts in mechanical resonance. For instance, when a membrane composed of a van der Waals magnet transitions from an (anti-)ferromagnetic to a paramagnetic state, a distinct change in strain leads to prominent softening or hardening of the resonant mode \cite{Siskins2020,CGTSiskins2021,Maurits2023,Houmes2023,Cheng2024}. When an external magnetic field is applied, the mechanical resonance frequency of suspended antiferromagnetic membranes exhibits abrupt shifts at antiferromagnetic-to-ferromagnetic spin-flip transitions, indicating a measurable exchange magnetostriction effect \cite{FaiMakJiang2020,Fei2024,Zhang2022,Li2022}. These transitions are also strain-tunable \cite{FaiMakJiang2020,Fei2024}, demonstrating exceptional spin-lattice coupling in these systems.   

Building on this framework, the exceptional sensitivity of nanomechanical resonators was exploited to detect antiferromagnetic spin texture dynamics in a standard laser interferometer setup (\textbf{Fig.~\ref{Fig:membranes}(a)}), revealing it through hysteretic magnetic responses and field-tunable mechanical anomalies \cite{Yousuf2025}. In this study, the mechanical detection of spin texture dynamics is realized by employing attometer-level lattice parameter change resolution with exceptional strain sensitivity in high-Q nanoelectromechanical resonators made of MnPS$_3$ (\textbf{Fig.~\ref{Fig:membranes}(b)}) in response to an external magnetic field due to magnetostriction (\textbf{Fig.~\ref{Fig:membranes}(c)}). The devices exhibit sharp, field-dependent shifts in resonance frequency and drops in quality factor near $4.6$ T and $7$ T, corresponding to a spin-flop transition and a magnetic domain reconfiguration, respectively (\textbf{Fig.~\ref{Fig:membranes}(d,e)}). These transitions are further attributed to rapid domain wall dynamics, supported by analytical magnetostriction modeling and atomistic spin dynamics simulations \cite{Yousuf2025}. Importantly, nonlinear mechanical responses of the same membranes reveal Duffing spring constants and nonlinear damping with abrupt field-tunable features, indicating a dynamic collective domain behaviour. A similar phenomenon was also observed in FePS$_3$ membranes, showing both temperature and magnetic field dependence of nonlinear parameters near the N\'{e}el temperature \cite{NonlinSiskins2025}. These results establish a robust and contactless platform for probing ultrafast spin texture phenomena in 2D magnets, opening new routes for spin-mechanical transduction affecting both linear and nonlinear dynamics of these systems.

Applicability of the nanomechanical method can be extended further by introducing angular resolution in measurements of resonance frequency, and thus strain, by covering a star-shaped array of rectangular cavities of high-aspect ratio between their length and width with 2D magnets \cite{Maurits2023, Houmes2023}. Such measurements have provided access to the anisotropic lattice deformations in antiferromagnetic 2D materials, such as FePS$_3$, CoPS$_3$, and NiPS$_3$ (\textbf{Fig.~\ref{Fig:membranes}(f)}), which can be directly related to their order parameters as a function of temperature (\textbf{Fig.~\ref{Fig:membranes}(g)}) \cite{Maurits2023}. Then, a change in critical behaviour with corresponding decrease of scaling exponent near the N\'{e}el temperature indicates the transition of magnetism from 3D to 2D as the magnetic materials are thinned down (\textbf{Fig.~\ref{Fig:membranes}(h)}) \cite{Maurits2023}. Experiments of the same type also showed prominent phase transition as well as the presence of quasi-1D magnetism in the magnetic sub-lattice of CrPS$_4$ \cite{Houmes2023}.

The outstanding sensitivity of mechanical resonance techniques has further enabled reconstruction of nontrivial magnetic phase diagrams in topological insulators. Nanomechanical experiments on MnBi$_2$Te$_4$ (MBST) revealed magnetic phase transitions from antiferromagnetic to canted antiferromagnetic to ferromagnetic phases (see \textbf{Fig.~\ref{Fig:membranes}(i)}) \cite{Liu2025}. This way, nanomechanical resonance measurements of MBST thin flakes can be used to directly probe magnetic phase transitions in an intrinsic antiferromagnetic topological insulator. By analysing the mechanical resonance signatures, a complex phase diagram of the temperature and field dependence of spin ordering in MBST can be reconstructed (\textbf{Fig.~\ref{Fig:membranes}(j)}) \cite{Liu2025}. These features are quantitatively modeled using an exchange magnetostriction framework, revealing field-dependent in-plane magnetostriction as the primary driver of the response \cite{Liu2025}. These experiments demonstrate potential for a scalable, noninvasive probe of spin-lattice interactions in layered quantum materials where spin, topological electronic properties, and mechanics are strongly entangled.

Arguably, the most long-coveted frontier involves coupling the nanomechanical motion of 2D materials to the quantum electronic landscape, where carbon nanotubes are currently still the more established system \cite{CNT1,CNT2}. Here, an important milestone was demonstrated by coupling a graphene nanoribbon membrane resonator and a single-electron transistor, allowing for sensing single-electron tunnelling \cite{Luo2017}. Another approach is to couple the mechanical system to quantum transport effects through capacitive softening of the membrane, by applying a finite gate voltage, introducing modulation of the membrane's chemical potential with its motion, and thus bringing quantum capacitance into the equation \cite{Chen2015, Manninen2022}. This effect renders the resonator sensitive to changes in the local electronic density of states \cite{Chen2015}. Notably, utilising this method, oscillations in the resonance frequency of graphene as a function of magnetic field have been shown to reflect the formation of Landau levels \cite{Chen2015}. Similarly, coupling to quantum capacitance allows for probing the De Haas–van Alphen effect in suspended graphene Corbino disk via shifts in the membrane's resonance frequency with an external magnetic field applied \cite{Manninen2022}. These mechanical readouts parallel to quantum Hall measurements can offer a noninvasive alternative to conventional transport techniques, with a specific benefit of access to the density of states of the low-dimensional system in a much simpler device geometry that requires only 1-2 contacts to the 2D material and a gate electrode at the bottom of the resonator’s cavity.

While nanomechanical sensing with 2D materials is rapidly redefining the boundaries of condensed matter characterisation, the most promising frontiers are still emerging. One key opportunity lies in the real-time sensing of domain dynamics in van der Waals magnets, where the spatiotemporal evolution of magnetic textures remains largely inaccessible to conventional probes \cite{Yousuf2025}. Similarly, topological phase transitions and the complex, correlated ground states found in twisted bilayer systems offer a compelling testbed for the next generation of resonance-based measurements employing coupling to quantum capacitance \cite{Chen2015, Zeng2025,Oh2021}. Here, although early studies have shown that quantum oscillations can modulate the resonance frequency of suspended membranes, the reverse remains an open question: can strain, engineered via nanomechanical methods, be used to tune and even control quantum states of matter? This feedback loop between mechanical and quantum degrees of freedom opens up new strategies in manipulating correlated states in TMDs for sensing applications. Looking ahead, there is also a growing interest in whether these techniques are sensitive enough to detect subtle thermodynamic signatures in elusive quantum phases in TMDs, such as in spin liquid \cite{MaasValero2021,Zhang2024} or excitonic insulator \cite{Lu2017} candidates. The potential to resolve minuscule entropy changes or energy anomalies through mechanical observables could add a new layer of insight into these strongly entangled systems. 

\section{Bio- and chemosensing with 2D materials}

2D materials offer intrinsically unique advantages for chemical and biological sensing applications. Their atomically thin structures and exceptionally high surface-to-volume ratios enable a large proportion of atoms to sense their environment. These properties result in a very low material loading, leading to an extremely economical resource and minimal environmental impact. Their mechanical flexibility, biocompatibility, and seamless integration with microfabrication techniques facilitate the creation of miniaturized, wearable, and on-chip sensor arrays with rapid response and recovery times \cite{Ma2021}. Facile surface functionalization and heterostructure manipulations impart superior selectivity toward specific gases, vapours, liquids, and their complex mixtures \cite{Kumar2021}. Complemented by high carrier mobility and tunable band gaps, 2D materials enable ultra-sensitive and low-power detection, particularly in field-effect transistor (FET)-based chemiresistive sensors \cite{Wang2019}.

\begin{figure*}[]
\includegraphics[width=14cm]{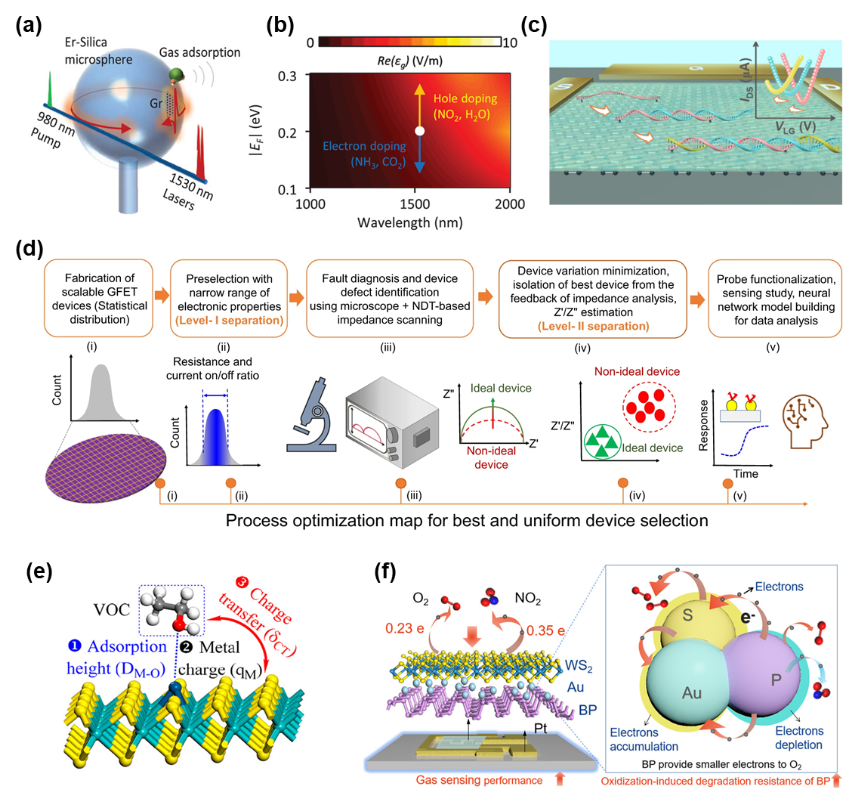}
\caption{\textbf{Bio- and chemosensing with 2D materials.} \textbf{(a)} Device composed of an erbium-doped silica microsphere with a diameter of 600 $\upmu$m as the microcavity and a p-doped graphene flake as the back scattering site. The device is pumped with a 980 nm laser and produces split modes with offset frequency dependent on the permittivity of the graphene flake, which varies with the types and concentrations of gas adsorption. \textbf{(b)} Effects of different adsorbed gas molecules on the Fermi level, and therefore the permittivity of graphene. The composition of a mixture of gases was calculated by linear deconvolution of their effects on multiple split modes. \textbf{(c)} A graphene FET probe coated by octadecyl alkylated trifluoromethyl-3-phenyldiazirine (C18-TPD) bonded via van der Waals forces to a DNA probe employed for sequential detection of DNA fragments through sequential hybridisation. \textbf{(d)} A schematic illustration of a nanofabrication approach that incorporates device preselection via impedance-based fault diagnosis to reduce variability and yield an optimised response. The testing and analysis workflow is illustrated in five steps (i–v). \textbf{(e)} A schematic illustration of the process of the absorption of volatile organic compounds (VOCs) in single-atom sensors (SASs) sites. Three descriptors: 1) metal charge (qM), 2) charge transfer between the anchored metal SASs-(supported by graphene/MoS$_2$/WSe$_2$) and VOCs ($\delta$CT), and 3) adsorption height (DM-O), represent the features of the different combinations of metal type, VOCs, and 2D substrates. \textbf{(f)} A 2D-0D-2D heterostructure toward high-performance room-temperature NO$_2$ sensing with black phosphorus (BP) (left) and the enhancement mechanism of the NO$_2$ sensors (right). The long-term stability of BP is achieved by the ultralow conduction band minimum position, the reduced adsorption energies of O$_2$, and the increased dissociation barrier energy of O$_{2}^{-}$ into 2O. Panels \textbf{a} and \textbf{b} are adapted from Ref.~[149], Wiley-VCH; panel \textbf{c} is adapted from Ref.~[152], American Chemical Society; panel \textbf{d} is adapted from Ref.~[159], Springer Nature; panel \textbf{e} is adapted from Ref.~[162], American Chemical Society; panel \textbf{f} is adapted from Ref.~[170], American Chemical Society.}
\label{fig:fig5chem}
\end{figure*}


As the breakthrough 2D material, the first proof-of-concept single molecule detection using graphene was reported in 2007 \cite{Schedin2007}. The measured step-like changes in the Hall resistivity of a three-layer graphene correspond to individual events of NO$_2$ molecule absorption and desorption. The high sensitivity was attributed to graphene’s exceptionally low electronic noise. Since then, there have been multiple examples of gas detection devices for polar gases like CO$_2$, NH$_3$, and NO$_2$ \cite{Dan2009, Kim2015, Kim2021}. These gases are of extreme importance in both industrial and biological applications, representing fundamental building blocks of both inorganic and organic materials, while being the main carriers of their respective elements in atmospheric environments (C and N). Their ultrasensitive detection is crucial for applications in environmental monitoring, medical diagnosis, industrial process control, and early warning of hazardous leaks. Recently, researchers have realized multispecies gas detection by measuring split modes of whispering-gallery-mode (WGM) generated by a graphene flake deposited on silica microspheres \cite{Guo2022} (\textbf{Fig.~\ref{fig:fig5chem}(a)}). By solving the linear equations related to multiple beat nodes, simultaneous detection with ppb-level sensitivities was reported for a mixture of CO$_2$, NH$_3$, NO$_2$, and H$_2$O with almost 100~\% recoverability within a few minutes (\textbf{Fig.~\ref{fig:fig5chem}(b)}). 

Intrinsically, graphene does not possess selective molecular interactions. Decoration and functionalisation of the graphene base layer opens a pathway for efficient and selective detection of higher molecular mass entities \cite{Kumar2021}. A surface-mounted metal–organic framework (SURMOF) on graphene field-effect transistors (FETs) showed exclusive sensitivity to ethanol against other alcohols. Furthermore, the robustness of graphene allows for \textit{in vitro} detection and discrimination of specific biologically relevant reactive species (e.g., •OH from other reactive oxygen species H$_{2}$O$_{2}$, •O$_{2}^{-}$)\cite{Wang2019}.

Due to their ultra-high sensitivity to adsorbed molecules, graphene-derived FETs are good candidates for next-generation label-free DNA sequencing (\textbf{Fig.~\ref{fig:fig5chem}(c)}). Graphene-based FETs were developed with a capability of detecting DNA nucleobases through conductivity measurements \cite{Dontschuk2015}. However, the technique was only applicable in specific conditions achieved in an ultrahigh vacuum chamber. The graphene FET technology was further modified for DNA nucleobase measurements applicable in solution \cite{Hwang2020}. Channels in crumpled graphene were employed as detection sites, proven more effective than flat graphene due to the electrical “hot spots” created by nano-deformation that leads to increased Debye length in the ionic solution, potentially allowing for speedy DNA sequencing. Further improvements in DNA and RNA fragment detection have been achieved, approaching practical applications by employing surface functionalisation with DNA probes \cite{Yang2025, Hwang2018}. They have shown greatly improved sensitivity with a limit of detection of 100~fM, and a dynamic range from 1~fM to 1~$\upmu$M. Furthermore, a similar technique for the detection of microRNA in biological fluid has shown prospective application for cancer detection from such tumor biomarkers \cite{Wang2024}.

Graphene oxide (GO) possesses oxygen functionalities on a graphene skeleton, improving biocompatibility and making it more suitable for \textit{in vivo} applications. GO-based fluorescent DNA aptasensor (based on aptamers - single-stranded oligonucleotides) was demonstrated for the diagnosis and treatment of liver cancer \cite{Ma2021}. Fluorescent DNA aptamers bound via hydrogen bonding, electrostatic, and $\pi$-$\pi$ stacking interactions are delivered to target sites, allowing both detection and targeted therapy, both \textit{in vitro} and \textit{in vivo}. This is proven by low cytotoxicity toward non-target cells.  Additionally, the oxygen-containing functional groups of GO provide numerous reactive sites for both covalent and non-covalent modifications, endowing GO-based sensors with greater adaptability and specificity toward a broad spectrum of analytes \cite{Lu2009, Chou2012, Dua2010, Robinson2008}. Arrays of rGO FETs functionalized with L-cysteine, thioglycolic acid, and anti-E.~coli antibodies were fabricated to act as selective absorbers \cite{Maity2023}. The devices realized ppb-level (10$^2$ cfu/ml level for bacteria) sensitive real-time monitoring of Pb$^{2+}$, Hg$^{2+}$, and E. coli in flowing water. Moreover, parallel fabrication of FETs was achieved on a full 4-inch wafer based on wet transfer of graphene oxide. Defective devices were screened out by non-destructive impedance spectroscopy and low‑frequency noise analyses. The device responses were calibrated by a machine learning algorithm to achieve coherence and accuracy across devices. This has showcased 2D materials as a promising path towards future intelligent and large-scale environmental monitoring networks (\textbf{Fig.~\ref{fig:fig5chem}(d)}).

Semiconducting 2D materials, such as TMDs, provide alternative strategies towards the development of chemosensing. 2H MoS$_2$-based implantable and absorbable biosensors can be employed for pressure and temperature measurements \textit{in vivo} \cite{Chen2018}. Such sensors exhibited low cytotoxicity when assessed \textit{in vitro}. Post-usage, the sensor dissolves in the body (in about 90 days) as shown by degradation in simulated body fluid (buffered phosphate solution, BPS), eventually transforming into relatively non-toxic products (Na$_2$S to Na$_2$SO$_4$ and MoO$_4^{2+}$). The low loading of material due to the 2D nature of the sensors further reduces the biological exposure.

The photoactive, semiconducting TMDs offer possibilities for self-powered sensing,  as exemplified by the development of gas sensors based on p-n type TMD-heterostructures \cite{Kim2020}. A WSe$_2$/WS$_2$ or WSe$_2$/MoS$_2$ junction served as a photovoltaic (PV) cell under daylight illumination and operated without external voltage. The response in short circuit current was measured when exposing the device to different concentrations of NO$_2$ (an electron donor) and NH$_3$ (an electron acceptor). With a 10-ppm detection limit and sub-second response, the possibility of self-sustained sensing platforms based on 2D semiconductors has been demonstrated.

Additionally, 2D materials can be functionalized to increase their sensitivity and stability as chemical sensors. Single-atom sensors (SASs) were constructed for the detection of volatile organic compounds (VOCs) at room temperature by functionalizing graphene, MoS$_2$, and WSe$_2$ with a series of monodispersed single atomic sites (Pt, Co, and Ru) (\textbf{Fig.~\ref{fig:fig5chem}(e)}) \cite{Liu2022}. The sensing performance of SASs was enhanced by 1-2 orders of magnitude in the relevant response of VOCs, and the mechanism was revealed by density functional theory (DFT). The introduction of single atoms modifies the charge dispersion and increases binding affinities for the adsorption of VOCs. Functionalizing 2D materials with single atoms constitutes a promising strategy to improve sensor performance and extend their applications.

An alternative approach towards self-powered chemosensors utilizes triboelectric effects \cite{Chang2020}. Polytetraﬂuoroethylene (PTFE) and latex-based triboelectric nanogenerator (TENG) were combined with organ-like Ti$_3$C$_2$T$_x$ MXene/MOF-derived Cu$_2$O nanohybrid \cite{Wang2021_2}. Ti$_3$C$_2$T$_x$ MXene fulfilled a double role of a standalone sensing material and the electrical transport interface in Cu$_2$O due to its metallic conductivity, which enables high response at the scale of power provided by the TENG. MXenes attracted manifold attention in the chemical sensing field. MXenes are early-transition metal carbides/carbonitrides and nitrides with a general formula M$_{n+1}$X$_n$T$_x$, where M represents transition metals (like Sc, Ti, V, Cr, Zr, Nb, Mo and Ta), X is C or N, and T represents surface terminal groups (like O, F, and OH)~\cite{MXene_gas_detector}. Their chemical sensing applications are based on special structures, abundant active sites, tunable surface chemistry, and outstanding stability. For example, 2D Ti$_3$C$_2$T$_x$ sheets were used as gas sensors to detect ethanol, methanol, acetone, and ammonia. Because of the active defective sites and interaction with surface functional groups, the sensing response can be very high, especially for ammonia (0.12/100 ppm)~\cite{MXene_gas_sensing}.

The diversity in the chemical and crystallographic structures provides unique advantages for sensing applications. This is exemplified by black phosphorus (BP), which is a layered allotrope of phosphorus with van der Waals interlayer bonds \cite{Antonatos2022}. BP is also sensitive to various molecules, such as NO$_2$, with comparable or better sensitivities than graphene and TMDs, as the adsorption energies of molecules are quite large \cite{Kou2014}. The first detection of NO$_2$ with BP was realized by creating BP-based FETs \cite{Abbas2015} and achieved excellent sensitivity down to 5 ppb. The relative conductance change agreed with the Langmuir isotherm model for molecular adsorption on a surface, indicating the sensing process was achieved by site binding and charge transfer. Meanwhile, the device presented good recovery to the original conductance after flushing with argon. The performance of BP-based sensors can be improved by heterostructure engineering. A 2D WS$_2$-Au-BP heterostructure was used for NO$_2$ detection \cite{Liang2024} and achieved a stable response of 2.11 to 100 ppb for 80 days (\textbf{Fig.~\ref{fig:fig5chem}(f)}). The enhancement of sensitivity was achieved by the promoted absorption energy and charge transfer in target molecules, as the formation of heterostructures reduced the potential energy barrier. Besides NO$_2$ detection, BP was also used in the detection of other chemical species, such as mercury \cite{Wang2025} and antibiotics \cite{Chen2023}. Other novel 2D materials, like SnS$_2$ \cite{Kumar2023}, MnPS$_3$ \cite{Kumar2020} are explored as potential candidates for NO$_2$ sensors. However, their stability and sensitivity are not well established when compared to traditional 2D materials. Te$@$Se heterostructures constitute an example of a stabilized system for NO$_2$ detection \cite{Cheng2025}. The device demonstrated an extremely high response of 622~\% to 1~ppm of NO$_2$ at room temperature, exhibiting an excellent stability over 90 days. Various strategies summarized herein highlight the versatility of 2D materials for application in chemical sensing.

\section*{Summary and outlook}

2D materials become increasingly important as building blocks for sensing applications. Their atomic thickness and mechanical flexibility, proximity effects in van der Waals heterostructures, and electrical control knobs in van der Waals devices constitute apparent benefits towards achieving high sensitivity for various physical quantities. The combinatorial approach to merging functionalities of different materials opens pathways towards multiproperty sensing. This can be realized at an unprecedented scale due to the emergence of a vast number of optoelectronic and spin states in 2D materials, which are highly tunable with unique 2D control knobs. We already know examples of realising sensing applications through a rational design of heterostructures combining multiple material systems. However, the parameter space arising from such an approach is immense, creating a need for the development of universal rules of 2D material assembly specifically tailored towards improved sensitivity, simplified detection techniques, and the understanding of underlying mechanisms. Multiple domains already encompass dedicated research on the implications of 2D properties towards sensing. However, it is apparent that the example practical realizations remain at an early stage.

\textbf{Optoelectronics:} The strong light–matter interaction in van der Waals materials, combined with their sensitivity to magnetic, electronic, and structural changes, offers powerful opportunities for non-invasive detection of quantum phases with unprecedented spatial and spectral resolution. This approach has been realized in multiple materials systems and can be extended to less explored compounds (e.g., XMenes), and specifically engineered heterostructures optimized for sensing methodology. This offers a means to probe exotic parameters and phase transitions. Beyond detection, optical control of quantum phases in these materials provides a non-contact, scalable route to engineer static and dynamic spin textures in 2D magnets for applications in magnetic storage, ultrafast memory, and reconfigurable spintronic devices. However, challenges persist, including the absence of clear optical signatures related to quantum phases in many materials, relatively weak coupling, extreme confinement of the exciton wavefunction limiting the range of interactions, interfacial charge transfer quenching optical responses, spectral congestion arising from multiple optically active levels, and inhomogeneities that obscure target states. Overcoming these hurdles will require scalable fabrication of defect-free heterostructures, enhanced coupling through photonic integration, and adaptive, high-fidelity detection schemes capable of isolating specific excitations and/or spin states.

\textbf{Quantum defects:} Despite rapid progress, quantum detection using emitters made of 2D materials still faces fundamental uncertainties. In most cases, the exact atomic structure of active defects remains a matter of speculation, and measurements are often based on ensembles rather than well-characterized single emitters. Fabrication and activation processes remain irreproducible, leading to high variability in their optical properties and sensing performance. Future efforts should aim at the deterministic creation of identified defects with uniform properties, combined with scalable integration into photonic and electronic architectures. Achieving homogeneous production on a semiconductor wafer scale would transform 2D quantum emitters from a collection of isolated demonstrations into a robust sensing technology.

\textbf{Scanning probes:} Merging 2D materials with scanning probe platforms is redefining what can be measured at the nanoscale. Graphene-coated AFM probes already surpass conventional coatings in electrical and chemical performance. Unique approaches such as QTM enable entirely new classes of experiments, most notably in situ twist-angle control for probing moiré physics on a single device. Different 2D material coatings can bring a variety of new sensing properties. E.g., defect-engineered hBN could be a promising candidate for scanning magnetometry, and QTM can map electronic and other interactions in diverse twisted structures. These advances point to a future where scanning probes are not just more sensitive or durable, but also quantum-enhanced, multifunctional instruments capable of mapping electronic, magnetic, mechanical, and thermal landscapes simultaneously. The main challenge currently lies in scaling fabrication methods and expanding the range of 2D materials for scanning probe microscopy beyond graphene.

\textbf{Nanomechanics:} Mechanical resonators already demonstrated a high applicability in sensing changes in material properties, making them a useful tool for detecting phase transitions - especially in systems where subtle effects such as magnetic spin textures or local spin reconstructions produce only weak signatures. The approach is particularly promising for studying those 2D magnets, where it can further help to uncover the spatial and temporal dynamics of domains or probe elusive ordering effects, like the formation of topological spin textures. Moreover, in twisted bilayer magnets, where competing interactions can give rise to coexisting magnetic phases, membrane-based sensing can provide a non-invasive route to disentangle their interplay. In the electronic domain of 2D materials, the coupling of nanomechanical motion to quantum phenomena can emerge as a powerful method to access the density of states with much simpler device geometries, while the prospect of using added strain to actively tune these quantum states opens up new experimental strategies for manipulating the strongly correlated systems.

\textbf{Bio- and chemosensing:} The major challenges in bio- and chemosensing arise from the synthesis of materials with suitable quality, bottlenecks in mass fabrication of individual devices, and the low accuracy of pinpoint functionalisation. Research into economical synthesis and functionalisation through scalable methods such as wet chemistry or roll-to-roll manufacturing, together with integration with existing semiconductor processing techniques (e.g., extreme ultraviolet lithography), should open the way to mass adoption. Once these obstacles are overcome, advanced 2D sensors will allow for extremely low-power, or self-powered, arrays of devices characterized by negligible mass and size. Potential technical and scientific applications include live chemical environment monitoring (e.g., contaminant detection in vacuum chambers and hydraulic systems), live monitoring of personal chemical environment (air, water, body fluids), and package-integrated food quality monitors.

\section*{Acknowledgements}
This project was supported by the Ministry of Education (Singapore) through the Research Centre of Excellence program (grant EDUN C‐33‐18‐279‐V12, I‐FIM). This research is supported by the Ministry of Education, Singapore, under its Academic Research Fund Tier 2 (MOE-T2EP50122-0012). This material is based upon work supported by the Air Force Office of Scientific Research and the Office of Naval Research Global under award number FA8655-21-1-7026. M. K. acknowledges funding from the MAT-GDT Program at A*STAR via the AME Programmatic Fund by the Agency for Science, Technology and Research under Grant No. M24N4b0034.

\bibliographystyle{unsrtnat}

\end{document}